\documentstyle[12pt,epsfig]{article}
\font\tenrm=cmr10

 1
\font\elevenrm=cmr10 scaled\magstep 1

\renewenvironment{thebibliography}[1]
 { \elevenrm
   \begin{list}{\arabic{enumi}.}
    {\usecounter{enumi}     \setlength{\parsep}{0pt}
     \setlength{\itemsep}{3pt} \settowidth{\labelwidth}{#1.}
     \sloppy
    }}{\end{list}}

\begin{document}
\title{\Large Restriction on the Neutron-Antineutron Oscillations
from the SNO Data on the Deuteron Stability 
\footnote{Presented at the 3-d International Workshop on Baryon and Lepton number violation
(BLV-2011), Gatlinburg, Tennessee, USA, Sept. 22-24 2011.}}

\author{Vladimir ~Kopeliovich$^a$\footnote{{\bf e-mail}: kopelio@inr.ru} 
and Irina ~Potashnikova$^b$\footnote{{\bf e-mail}: irina.potashnikova@usm.cl} 
\\
\small{\em a) Institute for Nuclear Research of RAS, Moscow 117312, Russia} \\
\small{\em b) Departamento de F\'{\i}sica, Centro de Estudios Subat\'omicos,}\\
\small{y Centro
Cient\'ifico - Tecnol\'ogico de Valpara\'iso,}\\\small{ Universidad T\'ecnica
Federico Santa Mar\'{\i}a, Casilla 110-V, Valpara\'iso, Chile}}
\maketitle
{\rightskip=2pc
 \leftskip=2pc
 \noindent}
{\rightskip=2pc
 \leftskip=2pc
\tenrm\baselineskip=11pt
\begin{abstract}
{ Restriction on the neutron-antineutron oscillation time in vacuum is obtained 
from latest SNO data on the deuteron stability, $\tau_D\,>\,3.01^.10^{31}$ years. 
Calculation performed within the quantum field theory based diagram technique
reproduces satisfactorily results of the potential approach previously developed.
The dependence of the obtained restriction on the total spin of the 
annihilating $N\bar N$ system
and the deuteron wave function modifications is discussed.}
\end{abstract}
 \noindent
\vglue 0.3cm}
\newpage 
\section{Introduction} 
 Searches for the baryon number violating processes predicted by Grand Unified Theories  
(GUT) remain to be an actual experimental task during many years.
The neutron-antineutron transition induced by the baryon number violating interaction
$(\Delta B=2)$ predicted within some variants of GUT has been discussed
in many papers since 1970 \cite{kuz}, see \cite{mm} --- \cite{alb}.
Experimental results of searches for such transition are available, in vacuum (reactor experiments
\cite{bal}, and references therein), in nucleus $^{16}O$ \cite{tak} and in $Fe$ nucleus \cite{berg}, 
in neutron magnetic trap \cite{serebrov}, see \cite{ckk,kkl}.

Here we derive a restriction on the neutron-antineutron transition time in vacuum using
new data on the deuteron stability obtained in Sudbury Neutrino Observatory (SNO) \cite{SNO}
and our former result on the suppression of $n-\bar n$ transition in deuterium \cite{kip}.
Restrictions on the $n-\bar n$ transition time in vacuum which follow from data
on nuclei stability have been obtained previously in a number of papers by different
methods, beginning with the potential approach of \cite{san,dr,dgr}, see \cite{aga94} -
\cite{faga}.

 To introduce notations, let us consider first the $n\bar n$ transition in vacuum
which is described by the baryon number violating interaction (see, e.g. \cite{mm,mik,lak}) 
$V= \mu_{n\bar n} \sigma_1/2$,  $\sigma_1$ being
the Pauli matrix. $\mu_{n\bar n}$ is the parameter which has the dimension of mass, to be predicted
by grand unified theories and to be defined experimentally \footnote{There is relation $\mu_{n\bar n}=2\delta m$
with the parameter $\delta m$ introduced in \cite{mm}. 
The neutron-antineutron oscillation time in vacuum is $\tau_{n\bar n}=1/\delta m=2/\mu_{n\bar n}$, see
also \cite{faga} and references in this paper.}. As usually, a point-like $n- \bar n$ coupling is assumed
here. The $n-\bar n$ state is described by the 2-component spinor $\Psi$,
lower component being the starting neutron, the upper one - the appearing antineutron.
The evolution equation is
$$ i{d\Psi \over dt} = (V_0+V) \Psi, \eqno (1) $$
where the diagonal matrix $V_0$ has  matrix elements  $V_0^{11}=V_0^{22}=m_N-i\gamma_n/2$ 
in the rest frame of the neutron ($m_N$ is the neutron (antineutron) mass, $\gamma_n$ -
the (anti)neutron normal weak interaction decay width, and we take $\gamma_{\bar n}=\gamma_n$, as it follows from
$CPT$-invariance of strong and weak interactions). Eq. $(1)$ has solution
$$ \Psi (t) = exp \left[-i\left(\mu_{n\bar n} t\,\sigma_1/2+V_0t\right)\right] \Psi_0 =
\left[cos{\mu_{n\bar n} t\over 2} -i \sigma_1 sin{\mu_{n\bar n} t\over 2}\right]exp(-iV_0t) \Psi_0, \eqno (2) $$
Here $\Psi_0$ is the starting wave function, e.g. for the neutron in the initial state $\Psi_0=(0,1)^T$.
In this case we have for the wave functions at arbitrary time $t$
$$\Psi (\bar n, t) = -i\,  sin{\mu_{n\bar n} t\over 2}exp(-iV_0t), \quad \Psi(n,t) = 
cos{\mu_{n\bar n} t\over 2} exp(-iV_0t) ,\eqno (3) $$
which describe oscillation $n-\bar n$:
In vacuum the neutron goes over into antineutron, also the discrete 
localized in space state, which can go over again to the neutron, so the oscillation
neutron to antineutron and back takes place indeed.

Since the parameter $\mu_{n\bar n} $ is small, the expansion of $sin$ and $cos$ can be made in Eq. $(3)$ at
not too large times. In this case the average (over the time $t^{obs} \ll 1/\mu_{n\bar n}$) change of the 
probability
of appearance of antineutron in vacuum is (for the sake of brevity we do not take into account the (anti)neutron
natural instability which has obvious consequences) is
$$ W(\bar n;t^{obs})/t^{obs}= |\Psi (\bar n, t^{obs})|^2/t^{obs} \simeq {\mu_{n\bar n}^2t^{obs}\over 4} \eqno (4) $$
which has, obviously, dimension of the width $\Gamma$.
From existing data obtained with free neutrons from reactor the oscillation time is greater than $0.86^.10^8 sec 
\simeq 2.7$ years \cite{bal}, therefore,
$$ \mu_{n\bar n} <  1.5^. 10^{-23} \; eV. \eqno (5) $$
\section{Suppression of $n-\bar n$ transitions in arbitrary nucleus}
 Recalculation of the quantity  $ \mu_{n\bar n}$ or $\tau_{n\bar n}$ from existing data on nuclei stability \cite{tak,berg,SNO}
is somewhat model dependent, and different authors obtained somewhat different results, within
about one order of magnitude, see
e.g. discussion in \cite{lak,bzk,faga}. 
In the case of nuclei the $n-\bar n$ line with the transition amplitude $\mu_{n\bar n}$ 
is the element of any amplitude describing the nucleus decay $A \to (A-2) +\, mesons$, where
$(A-2)$ denotes a nucleus or some system of baryons with baryonic number $A-2$.
The decay probability is therefore proportional to $\mu_{n\bar n}^2$,
and we can write by dimension arguments
$$ \Gamma (A \to (A-2)\, +\, mesons) \sim {\mu_{n\bar n}^2 \over m_0},\eqno (6) $$
where $m_0$ is some energy (mass) scale.

 It was argued in \cite{kip} that $m_0$ is of the order of normal hadronic 
or nuclear scale, $m_0 \sim m_{hadr}\sim (10  - 100)\,MeV$.
The dimensionless suppression factor is therefore

$$ F_S = {\mu_{n\bar n} \over m_0} < 10^{-30},\eqno (7) $$
We can obtain the same result from the above vacuum formula $(4)$, 
if we take the observation 
time $t^{obs} \sim 1/m_{hadr}$.

The physical reason of such suppression is quite transparent and has been discussed 
in the literature long ago (see e.g. \cite{mm,mik,aga}):
it is the localization of the neutron inside the nucleus, whereas no localization
takes place in the vacuum case.
\section{The deuteron decay probability}
 The case of the deuteron, which is most simple and 
instructive, can be treated using
the standard diagram technique. Such technique or its modifications 
have been used in \cite{vin,lak} and \cite{bzk}
\footnote{Later the author of \cite{vin} tried to develope a field theory motivated 
approch to this problem which was criticized by scientific community 
\cite{aga94,mik,bzk,kip}.}.

The point is that in this case there is no final state containing antineutron ---
it could be only the $p\bar n$ state, by the charge conservation. But this state is 
forbidden by energy conservation, since the deuteron mass is smaller than the sum of
masses of the proton and antineutron.
Therefore, if the $n-\bar n$ transition took place within the deuteron, the final
state could be only some amount of mesons.
\begin{figure}[h]
\label{deut}
\setlength{\unitlength}{.8cm}
\begin{flushleft}
\begin{picture}(10,6)
\put(3,2){\line(1,0){8.}}
\put(3,2.5){\line(1,0){2.}}

\put(5,2.25){\circle{0.5}}
\put(4.5,1.4){$g_{Dnp}$}
\put(11.,2.25){\circle*{0.5}}
\put(9.5,1.2){$T(\bar n p \to mes.)$}

\put(11.,2.2){\line(2,-1.){1.7}}
\put(11.,2.2){\line(5,-1){2.1}}
\put(11.,2.2){\line(5,1){2.1}}
\put(11.,2.2){\line(1,0){2.2}}
\put(11.,2.2){\line(2,1.){1.7}}

\put(8,5.5){\circle*{0.2}}
\put(7.4,5.9){$\mu_{n\bar n}/2$}

\put(2.6,2){$D$}
\put(7.8,1.6){$p$}
\put(13.3,2.2){$mesons$}
\put(6.3,4.2){$n$}
\put(9.4,4.2){$\bar n$}

\put(5.,2.5){\line(1,1){3}}
\put(8.,5.5){\line(1,-1){3}}

\end{picture}
\caption{The Feynman diagram describing $n - \bar n$ transition in the deuteron with
subsequent annihilation of antineutron and proton to mesons.}
\end{flushleft}
\end{figure}
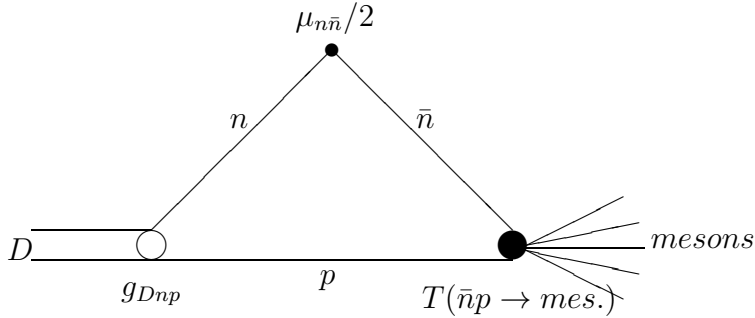
The amplitude of the process is described by the diagrams of the type shown in
Fig.1 and is equal to
$$ T(D\to mesons) = 2ig_{Dnp}m_N^2\mu_{n\bar n} \int {T(\bar n p \to mesons)\over (p^2 -m_N^2)[(d-p)^2-m_N^2]^2}
{d^4p\over (2\pi)^4}.  \eqno (8)$$
Here $p$ and $d$ are the 4-momenta of the virtual proton and deuteron. 
The constant $g_{Dnp}$ is normalized by the condition 
$${g^2_{Dnp}\over 16\pi } = {\kappa\over m_N} = \sqrt{{\epsilon_d\over m_N}}\simeq 0.049, \eqno (9)$$
which follows, e.g. from the deuteron charge formfactor normalization $F_D^{ch}(t=0)=1$,
see discussion and references in  \cite{kip}.
$\kappa =\sqrt{m_N\epsilon_D}\simeq 46\,Mev$, $\epsilon_D \simeq 2.23\,MeV$ being the binding energy of the deuteron.
In the nonrelativistic reduction of Feynman diagram (Fig. 1) we should write
for the vertex $D \to np$ $2m_Ng_{Dnp}$ and $m_N \mu_{n\bar n}$ for the $n \to \bar n$
transition amplitude, to ensure the correct dimension of the whole amplitude
(here we correct some inaccuracy of our former consideration in\cite{kip}).

Presence of the second order pole in the energy variable of intermediate nucleon is a
characteristic feature of diagrams describing the $n - \bar n$ transition in nuclei,
as discussed in \cite{kip}. This does not lead, however, to any dramatic consequences,
and the integration over internal 4-momentum $d^4p$ in $(8)$ can be made easily taking into account the nearest
singularities in the energy $p_0=E$, in the nonrelativistic approximation for nucleons. 
The integral over $d^3p$ converges at small $p\sim \kappa$ which corresponds to large 
distances between nucleons in the deuteron, $r\sim 1/\kappa$.
By this reason the annihilation amplitude can be taken out of the integration on the mass shell in some average point,
and we obtain the approximate equality \cite{kip}
$$ T(D\to mesons) = 2g_{Dnp}m_N^2\mu_{n\bar n} I_{DNN} T(\bar n p \to mesons) \eqno (10)$$
with the integral
$$I_{DNN}= {i\over (2\pi)^4}\int {d^4p \over (p^2-m_N^2)[(d-p)^2-m_N^2]^2} \simeq  $$
$$ \simeq {i\over (2\pi)^4 (2m)^3 }\int {d^4p\over \left[p_0-m_N-\vec p^2/(2m_N) +i\delta\right]
\left[m_D -m_N-p_0-\vec p^2/(2m_N) -i\delta\right]^2} = $$ 
$$=\int {d^3 p \over (2\pi)^3 8m_N[\kappa^2+\vec p^2]^2}={1\over 64\pi m_N\kappa}, \eqno (11) $$
There is close connection between the amplitude of the $n-\bar n$ transition in deuteron
and the deuteron charge formfactor at zero momentum transfer, which contains
the integral $I_{DNN}$ as well  \cite{kip},

Using the standard technique, we obtain for the decay width (probability):
$$ \Gamma (D\to mesons) \simeq \mu_{n\bar n}^2 g_{Dnp}^2 I_{DNN}^2 m_N^3
\int |T(\bar n p \to mesons)|^2  d\Phi (mesons), \eqno (12) $$
$\Phi (mesons)$ is the final states phase space.
Our final result for the width of the deuteron decay into mesons is
$$\Gamma_{D\to mesons} \simeq {\mu_{n\bar n}^2\over 64\pi\kappa} m_N^2\left[v_0\sigma^{ann,S=1}(\bar n p)\right]
_{v_0\to 0}\simeq
{\mu_{n\bar n}^2\over 32\pi\kappa} m_N \left[p_{c.m.}\sigma^{ann,S=1}_{\bar n p}\right]_{p_{c.m.}\to 0}, \eqno (13)$$
where $p_{c.m.}$ is the (anti)nucleon momentum in the center of mass system.
This result is close to that obtained in \cite{vin} and in \cite{lak} where it was
obtained using the induced $\bar n p$ wave function. \footnote{The result 
of \cite{lak} given by Eq. (17) can be rewritten in our notations as
$$ \Gamma_{D\to mesons} \simeq 0.01 \mu_{n\bar n}^2 {m_N^2\over \kappa} \left[v_0\sigma^{ann}{\bar n p}\right]
_{v_0\to 0},\qquad \qquad (17')$$
which is abot twice greater than our result.}.

The annihilation cross section of the antineutron with velocity $v_0$ on the proton at rest equals
$$\sigma^{ann}_{\bar n p}= \sigma (\bar n p \to mesons) = {1\over 4m_N^2v_0}\int |T(\bar n p \to mesons)|^2  d\Phi(mesons).\eqno (14) $$
According to PDG at small $v_0$, roughly, $\left[v_0 \bar\sigma^{ann}_{\bar n p}\right]_{v_0\to 0}\simeq (50 - 55) mb 
\simeq (130 - 140) \,GeV^{-2}$ where averaging over antineutron and proton spin variables is assumed.
 We obtain from Eq. $(13)$ for the deuteron life time 
$$\tau_D = {4\over \mu_{n\bar n}^2} {8\pi\kappa\over 
 m_N \left[p_{c.m.}\sigma^{ann, S=1}_{\bar n p}\right]_{p_{c.m.}\to 0}}
= \tau_{n\bar n}^2 {16\pi\kappa\over 
 m_N^2 \left[v_0\sigma^{ann, S=1}_{\bar n p}\right]_{v_0\to 0}}, \eqno (15) $$
$v_0$ is the antiproton velocity in the laboratory frame where proton is at rest.

If we define the suppression factor (dimensional) $R_D$ as usually,
$$ \tau_D = \tau_{n\bar n}^2 R_D, \eqno (16) $$
with $\tau_{n\bar n} = 2/\mu_{n\bar n}$, then we have from Eq.(15)  that
$$ R_D = 
{16\pi\kappa \over m_N^2  \left[v_0\bar \sigma^{ann,S=1}_{\bar n p}\right]_{v_0\to 0}} \eqno (17))$$
Quite naturally suppression increases with increasing binding energy of the
deuteron $\epsilon_D$, or $\kappa$. Numerically we have $R_D \simeq 2.94^. 10^{22} sec^{-1}$
if we neglect possible dependence of the annihilation cross section on the total spin
of the $\bar n - p$ system.
This is in agreement with results of the papers \cite{dgr} where $R_D$ was found to be in
the interval between $2.4^.10^{22}$ and $2.75^.10^{22}\, sec^{-1}$ for different variants of
the potential model.

To get an idea what happens for heavier nuclei, we can use the formula $(17)$ with
greater value of $\epsilon$, about $\sim 8\,MeV$, the binding energy per one nucleon 
in heavy nucleus. This leads to the value
$R_A\sim 5.5\, 10^{22}\, sec^{22}\,sec^{-1}$, in qualitative agreement with \cite{dgr}.
We obtain same estimate if we replace the deuteron size $r_D=1/\sqrt{\epsilon m_N}$ by
average internucleon distance in heavy nucleus, $r_{inter-N}\sim 2\,Fm$.
\section{The role of the total spin dependence of the $\bar n p$ annihilation cross section}
 This result can be refined taking into account spin dependence of the $N\bar N$
annihilation amplitudes and going beyond the zero range approximation for the deuteron
wave function.

Within the deuteron neutron and proton are in triplet state, and after $n\to \bar n$ 
transition antineutron and proton remain in triplet state, therefore, only part of
the total annihilation cross section corresponding to triplet state works in our case.

 We can write for the annihilation cross section averaged over spin states 
$$ \bar \sigma^{ann}_{\bar n p} ={1\over 4}(3\sigma^{ann,S=1}_{\bar n p} + \sigma^{ann,S=0}_{\bar n p})
= {3r+1\over 4r}\sigma^{ann,S=1}_{\bar n p} \eqno (18) $$
where $\sigma^{ann,S=1}_{\bar n p}$ and $\sigma^{ann,S=0}_{\bar n p}$ are the triplet and singlet
annihilation cross sections, 
$r=\sigma^{ann, S=1}_{\bar n p}/\sigma^{ann,S=0}_{\bar n p}$,
or $\sigma^{ann,S=1}_{\bar n p}= 4r\bar \sigma^{ann}_{\bar n p}/(3r+1). $

From these results and definitions we obtain the final formula
$$ \tau_{n\bar n} = \left\{\tau_D\frac{r\,m^2_N\left[v_0\sigma^{ann}_{\bar n p}\right]_{v_0\to 0}}   
{4\pi\kappa (3r+1)} \right\}^{1/2}. \eqno (19) $$
The supression factor is
$$ R_D = 
{4(3r+1)\pi\kappa \over r\,m_N^2  \left[v_0\bar \sigma^{ann}_{\bar n p}\right]_{v_0\to 0}}. \eqno (20)$$
Numerically $R_D \simeq 2.94 ^. 10^{22} (3r+1)/4r \; sec^{-1} $, and from new SNO data \cite{SNO}
$\tau_D\,>\,3.01^.10^{31}$ years, we obtain 
$$\tau_{n\bar n}>  1.8^.10^8\left[4r/(3r+1)\right]^{1/2}\,sec. \eqno (21)$$
At $r=1$ we recover our former result for the spinless case, if $r\gg 1$ we would obtain
$$\tau_{n\bar n}>  2.1^.10^8 \,sec \eqno (22)$$
In the paper \cite{kk} it was obtained from the combined analysis of the $\bar p p$ atom data
and results of OBELIX scattering experiments at LEAR that for the antiproton-proton 
interactions
$r_{\bar p p}\simeq 0.42$. The $\bar p p$ state is a mixture of isoscalar and isovector, 
whereas $\bar n p$ is pure isovector, but
if we take the same ratio for the $\bar n p$ interaction, we obtain from the 
SNO data \cite{SNO} that $\tau_{n\bar n}>  1.55^.10^8 \,sec. $
\section{Possible deuteron wafe function modifications; \\ conclusions}
 To go beyond the zero range approximation for the deuteron wave function we can use e.g.
the Hulthen deuteron wave function in the form
$$\Psi_{D,H}(\vec r) = {1 \over \sqrt{2\pi}} {\sqrt{\alpha\kappa(\alpha + \kappa)}\over
\alpha - \kappa} {\left[\exp(-\kappa r) - \exp (-\alpha r)\right]\over r},  \eqno(23) $$
where $\alpha \simeq 270\,Mev$ is the Hulthen parameter \cite{wong}.
In the limit $\alpha \rightarrow \infty$ we obtain the zero range deuteron wave function
$\Psi_D=\sqrt{\kappa/2\pi} exp(-\kappa r)/r$.

The additional formfactor appears in the integrand of the amplitude in Eq. (11) when we are 
using the Hulthen wave function. Instead of the integral
$$ \int {d^3p\over [\kappa^2+\vec p^2]^2}={\pi^2\over \kappa} $$
we obtain now 
$$\sqrt{\alpha(\alpha+\kappa)^3} \int {d^3p\over [\kappa^2+\vec p^2]^2\left(\alpha^2+\vec p^2\right)} =
{\pi^2\over \kappa} \sqrt{\alpha\over \alpha+\kappa } . \eqno (24)$$
So, using more realistic deuteron wave function leads to additional factor
$\sqrt{\alpha\over \alpha+\kappa } \simeq 0.92 $ which does not change substantionally
our results: the final estimate for $\tau_{n\bar n}$ decreases by about $4\%$.
This means also that possible modifications of the deuteron wave function at small internucleon 
distancies will not change our results considerably.

To conclude, the results presented here are in satisfactory agreement with previously
obtained results of the potential approach by Dover, Gal and Richard \cite{dgr}, 
and also in crude agreement with \cite{lak,vin}. Our method opens a way to study 
possible relativistic effects and corrections.
Restriction on the $n-\bar n$ transition time in vacuum obtained from the $SNO$ data
\cite{SNO} is comparable with that from data on heavier nuclei stability. The best
restriction which comes from data on oxygen stability, is $\tau_{n\bar n} >3.3^.10^8\, sec$ \cite{faga}.
\section{ Acknowledgements}
We are thankful to Yu.A.Kamyshkov and B.Z.Kopeliovich who encouraged us to make these 
estimates, and to B.O.Kerbikov for useful discussion of the spin dependence of $\bar N N$
annihilation probability.
This work has been supported by Fondecyt (Chile), grant number 1090236.
\\

\end{document}